\documentclass[3p,times,twocolumn]{elsarticle}

\usepackage{lineno,hyperref}
\usepackage{subfigure}
\modulolinenumbers[5]

\usepackage{amssymb}
\usepackage{graphicx}
\usepackage{subfigure}

\journal{Journal of Alloys and Compounds}

\begin{document}

\begin{frontmatter}



\title{CuAlTe$_2$: A promising bulk thermoelectric material}


\author{Vijay Kumar Gudelli$^a$, V. Kanchana$^{a,*}$, G. Vaitheeswaran$^b$}

\address{$^a$Department of Physics, Indian Institute of Technology Hyderabad, Ordnance Factory Estate, Yeddumailaram-502 205, Telangana, India\\
         $^b$Advanced Centre of Research in High Energy Materials (ACRHEM), University of Hyderabad, Prof. C. R. Rao Road, Gachibowli, Hyderabad-500 046, Telangana, India\\
         $^*$Corresponding Author E-mail: kanchana@iith.ac.in}

\begin{abstract}
Transport properties of Cu-based chalcopyrite materials are presented using the full potential linear augmented plane wave method and Boltzmann Semi-classical theory. All the studied compounds appear to be  direct band gap semiconductors evaluated based on the Tran-Blaha modified Becke-Johnson potential. The heavy and light band combination found near the valence band maximum (VBM) drive these materials to possess good thermoelectric properties. Among the studied compounds, CuAlTe$_2$ is found to be more promising, in comparison with CuGaTe$_2$, which is reported to be an efficient thermoelectric material with appreciable figure of merit. Another interesting fact about CuAlTe$_2$ is the comparable thermoelectric properties possessed by both $n-$ type and $p-$ type carriers, which might attract good device applications and are explained in detail using the electronic structure calculations.

\end{abstract}


\end{frontmatter}


\section{Introduction}
\label{intro}
The Cu-based chalcopyrite-type (CCP) materials are well-known semiconductors with energy band gap ranging from 1-3 eV \cite{Yu,Xiao}. The band gap range of these materials includes almost all the frequencies of the solar spectrum. Similarly, the semiconducting nature of these materials attracts the present researcher, as these materials find promising applications in solar cells, detectors, non- linear optical devices and many more. The important application of the chalcopyrite-type compounds includes the energy conversion and serve as photovoltaic and thermoelectric energy converters. Most of the previous studies on CCP systems have concentrated more on the optical properties, which has laid the path for highly efficient photovoltaic and solar energy applications. For example, Cu(In,Ga)Se$_2$ based system is a remarkable solar cell material with an energy conversion efficiency of 20.3$\%$ \cite{Jackson,Contreras,Verma}. Apart from the photovoltaic and solar energy cell applications, the CCP also possess good thermoelectric (TE) properties. In particular, CuGaTe$_2$ showed a dimensionless figure of merit of ZT = 1.4 from experiment \cite{Plirdpring} and theoretical studies \cite{VKG1,Daifeng} reported ZT to be 1.69 at a maximum temperature of 950 K, which is in comparison with the state-of-the-art Bi$_2$Te$_3$ \cite{Poudel} and PbTe-based \cite{Hsu} alloys. Other doping studies are also initiated in order to improve the performance of the CuGaTe$_2$ \cite{Jiaolin,Aikebaier}. This motivates the present study of analysing the iso-structural compounds CuAlS$_2$, CuAlSe$_2$, CuAlTe$_2$, CuGaS$_2$ and CuGaSe$_2$ to explore their thermoelectric properties.
\par
Thermoelectric (TE) materials with potential applications in power generation and refrigeration have been a thrust area of research for the past few years. TE materials can convert waste heat into electric power and therefore can be of immense use in meeting the present status of energy crisis and environmental contamination \cite{Majumdar,Bell,Snyder,Dress}. The performance of a TE material is represented by the dimensionless figure of merit, ZT, given by $ZT = S^2$ $\sigma T$ $/ \kappa$, where $S$, $\sigma$, $\kappa$ and T are the thermopower, the electrical conductivity, the thermal conductivity, and the absolute temperature, respectively. $\kappa$ includes both the electronic $\kappa_e$, and the lattice contributions $\kappa_l$, i.e., $\kappa=\kappa_e + \kappa_l$. To have a good figure of merit the materials should have a high thermopower like doped semiconductors/insulators and metal like electrical conductivity. At the same time, the material should have low thermal conductivity. Achieving all these requirements together in a particular material is a challenge in the present research field. Despite these conflicting requirements, remarkable progress has been realised in recent years exploring different classes of materials for better TE performance \cite{Snyder,Dress,Georg,singh,Abdeljalil,Jiong,andrew,Wang,Jovovic,David,YuLi,khu,Lijun,Parker,Khuong,Mun}. However, having good figure of merit alone will not serve the energy demand, while there are other materials properties to be considered.  The materials should be structurally stable in the operating temperature range and have a high melting point as far as waste heat recovery is concerned. In addition, it is mandatory that the elements comprising the materials should be abundant and cheap. Our earlier studied compound CuGaTe$_2$ possess these properties and we intend to explore the iso-structural compounds CuAlS$_2$, CuAlSe$_2$, CuAlTe$_2$, CuGaS$_2$ and CuGaSe$_2$, which might also show good TE properties.
	To understand the TE properties of these compounds we have performed a systematic first principle calculations based on the density functional theory. The rest of the paper is organised as follows: Section 2  describes  the methodology, and section 3 presents the results and discussions. Conclusions are given in section 4.

\section{Methodology}
\label{sec:1}
Total energy calculations based on first principle density functional theory (DFT) were performed using the full-potential linear augmented plane wave (FP-LAPW) method as implemented in WIEN2k \cite{Blaha}. The total energies are obtained by solving the Kohn-Sham equations self-consistently within the Generalised Gradient Approximation (GGA) of Perdew-Burke-Ernzerhof (PBE) \cite{Perdew} and the electronic band structures are calculated using the optimised parameters. The self-consistent calculations included  spin-orbit coupling. As it is well known that the calculations using standard local-density (LDA) or Generalised Gradient Approximation (GGA) schemes for the exchange-correlation potential underestimate the band gaps of semiconductors, we have used the Tran-Blaha modified Becke-Johnson \cite{Becke} potential  (TB-mBJ) \cite{Tran1}. We have used plane waves with a cutoff of $R_{MT}$K$_{max}$ = 7, where R$_{MT}$ is the smallest of all atomic sphere radii, and K$_{max}$ is the plane wave cut-off. The maximum  value for the wave function expansion inside the atomic spheres was restricted to $l_{max} = 10$.
For k-space integrations, a $20\times20\times20$ k-mesh was used for CuXCh$_2$ in the Monkhorst-Pack scheme \cite{Monkhorst}, resulting in 641 k-points in the irreducible part of the Brillouin zones for all the compounds, respectively. The crystal structure of CuXCh$_2$ is tetragonal with space group $I\bar{4}2d$  (no. 122). All the calculations were performed with an energy convergence criterion of $10^{-6}$ Ry per formula unit. The carrier concentration (p for holes and n for electrons) and temperature ($T$) dependent thermoelectric properties like thermopower (S), electrical conductivity ($\frac{\sigma}{\tau}$) were computed using BOLTZTRAP \cite{Madsen} code, within the Rigid Band Approximation (RBA) \cite{Scheidemantel,Jodin,Chaput} and the constant scattering time ($\tau$) approximation (CSTA). According to RBA approximation, doping a system does not alter its band structure but varies only the chemical potential, and it is a good approximation for doped semiconductors to calculate the transport properties theoretically when doping level is not very high \cite{Jodin,Chaput,Bilc,Ziman,Nag}. However certain dopant can drastically modify the nature of electronic structure near the gap giving rise to resonant states \cite{Bilc_04,Ahmad} in which case, RBA can fail \cite{MSlee}. According to CSTA, the scattering time of the electron is taken to be independent of energy and depends only on concentration and temperature. The detailed explanation about the CSTA can be found elsewhere \cite{singh,Khuong,aggate2}. It is evident that CSTA has been quite successful in the past in predicting the thermoelectric properties of many materials \cite{Georg,Lijun,Parker,Khuong,DJS}.

\section{Results and Discussion}
\subsection{Band structure and Density of States of CuAlCh$_2$ and CuGaCh$_2$}
The structural properties obtained with optimised lattice parameters and internal positions are given in Table-1, along with the available experimental and other theoretical results. From Table-1, it is evident that the optimised parameters are in good agreement with the available experiment and other theoretical calculations. The band structure of all the compounds are computed using TB-mBJ exchange correlation functional. 
The overall profile of the calculated band structures of all the investigated compounds are very similar to each other, and the band structure of CuAlS$_2$ along the high symmetry directions in the tetragonal Brillouin zone alone is shown in Fig. 1(a). The band structure of CuAlS$_2$ reveals a direct band gap at the center of the Brillouin zone i.e. at $\Gamma$, which is the same in all the compounds of the present study. The calculated band gaps are tabulated in Table-2 and it is evident that there exist a difference between the calculated and experimental band gaps. But it is to be noted that our band gaps are slightly lower compared with the band gaps of ref. \cite{Raja} which are also calculated using TB-mBJ functional. The possible reason for this might be due to the inclusion of spin-orbit coupling in the present work. The band found just below the valence band maxima (VBM) arises from the $Cu-d$ and chalcogen-$p$ states, and below this lies a band of chalcogen-$p$ and $Al(Ga)-d$ character. The conduction band minima (CBM) is mostly the mixer of chalcogen-$p$ states with $Al(Ga)$-$p$ states. From Fig. 1(a), the vicinity of VBM is composed of nearly degenerate bands (we call them as upper VBM(I) and lower VBM(II)) with different energies along the high symmetry direction $\Gamma-N$, while they are degenerate along $\Gamma-Z$. The inset of Fig. 1(a) show the upper and lower VBM. In the case of CBM there are no such degenerate bands. To explore further about the carriers at VBM and CBM, we have calculated the effective mass of the bands along $\Gamma-N$ and $\Gamma-Z$ directions and are shown in Table-3. We note a significant difference in the mass of the carriers of the upper and lower VBM in all the investigated compounds. The difference in the effective mass of VBM(I) and VBM(II) along $\Gamma-N$ is because of the variation in the dispersion of the bands at the VBM and will serve as mixer of light and heavy bands, which will often be favourable for thermoelectric performance. In the case of the CBM, we find the bands with low effective mass, which might favour high electrical conductivity. The effective mass of the carriers in VBM along both $\Gamma-N$ and $\Gamma-Z$ directions is found to decrease from $S$ to $Te$ in the case of CuAlCh$_2$, and also in the case of CuGaCh$_2$ (see Table-3). This nature is because of the increase in the dispersion of the bands from $S$ to $Te$, which can be seen clearly in the density of states. But in case of CBM along both $\Gamma-N$ and $\Gamma-Z$ directions the trend is different, the effective mass of the carriers is found to decrease from $S$ to $Se$ (in both $Al$ and $Ga-$ compounds) then increase from $Se$ to $Te$ in CuAlCh$_2$. 

\par The Density of States (DOS) of CuAlS$_2$ is shown in Fig. 1(b) along with the specific contribution of the constituent atoms present in the compound. From Fig. 1(b), it is evident that the VBM is composed of $Cu-d$ and chalcogen-$p$ states, whereas in CBM it is composed of chalcogen-$p$ states with $Al(Ga)$-$p$ orbital as discussed earlier in the band structure. The heavy mass carriers at the VBM are responsible for the strong increase in the DOS at the VBM. However, the DOS also rises steeply above the CBM, albeit not as distinctly as around the VBM. We have also shown the DOS of all the compounds in Fig. 2, which shows that the sharp increase in the DOS near the VBM is found to decrease from $S$ to $Te$ in CuAlCh$_2$, which implies a decrease in the effective mass from $S$ to $Te$ in case of CuAlCh$_2$. A similar situation is observed for CuGaS$_2$ and CuGaSe$_2$. In case of the CBM there is a steep increase in the DOS of CuAlTe$_2$ compared to CuAlS$_2$ and CuAlSe$_2$ which is because of the increase in the effective mass of $Te$ compared to $S$ and $Se$ in CuAlCh$_2$. The combination of light and heavy bands favour better thermoelectric performance, the heavy bands usually contribute to a high thermopower while the lighter bands offer an advantage of high mobility. This will be discussed in the following section.

\subsection{Thermoelectric properties of CuAlCh$_2$ and CuGaCh$_2$}
The motivation that the mixed nature of the bands might offer a good thermoelectric performance, further lead us to calculate the thermoelectric properties of the CCP. The carrier concentration and temperature dependent TE properties of CuAlCh$_2$, CuGaS$_2$ and CuGaSe$_2$ are calculated using BoltzTrap code within the limit of RBA and CSTA as mentioned in section-2. The thermoelectric properties such as thermopower ($S$ in $\mu$ $V/K$), electrical conductivity scaled by relaxation time ($\sigma/\tau$ in $(\Omega m s)^{-1}$) and power factor scaled by relaxation time ($S^2 \sigma/\tau$ in $W/m K^2 s$) are calculated for both electron ($n_e$) and hole ($n_h$) concentrations at different temperatures. As the investigated systems are of tetragonal structure, we have also studied the effect of anisotropy in the thermoelectric properties along both $a$ and $c$-directions. The calculated thermoelectric properties of CuAlS$_2$ as a function of both the carrier concentration (electrons on left and holes on right panel) at different temperatures are presented in Fig. 3. From these figures, it is quite evident that there is no signature of bipolar conduction which might be because of the higher band gaps ($>1 eV$) of all the compounds. We intend to look for materials with high efficiency and capable of finding application at high temperature, which lead us to compare the thermoelectric properties of all the compounds at 700 $K$ and are shown in Fig. 4. Thermoelectric properties of CuAlS$_2$ are shown in Fig. 3, and it is evident from the figure that the thermopower is higher in the case of holes (above +700 $\mu$ $V/K$) compared to electrons (above +600 $\mu$ $V/K$). We also found that there is small anisotropy in the thermopower values in the case of electron, while $\sigma/\tau$ is found to be more anisotropic in the case of holes compared to electrons, with the value being higher for electrons compared to holes. Although the thermopower is high for the hole carriers, the power factor of CuAlS$_2$ is found to be little higher in case of electrons compared to holes, which is because of the higher $\sigma/\tau$ values possessed by them. Similar situation is observed in other compounds also. From Fig. 4, it is clear that the thermopower of holes is found to be more compared to electrons in all the compounds and the same is also found to decrease from $S$ to $Te$ in the case of CuAlCh$_2$ for holes, whereas in the case of electrons it decreases from $S$ to $Se$ and then increases in $Te$, which might be because of the heavy effective mass of the electrons in CuAlTe$_2$. A similar trend is also seen in CuGaCh$_2$  \cite{VKG1}. 
The anisotropy is almost similar for both electrons and holes and it is found to increase from $S$ to $Te$ in both CuAlCh$_2$ and CuGaCh$_2$. 
The values of $\sigma/\tau$ is found to be more along $a$-axis compared to the $c$-axis. 
Among the CuAlCh$_2$ compounds the power factor of CuAlTe$_2$ is found to be higher compared to the other two chalcogens with electrons as the charge carrier along $a-$axis.
All the investigated compounds have shown good TE properties for electrons. 
The low effective mass for electrons have shown high electrical conductivity leading to high power factor. Hence it implies that the carriers with low effective mass may give a better performance for thermoelectric materials \cite{Pei}.

\subsection{Comparison of CuAlTe$_2$ with CuGaTe$_2$}
Earlier studies on CuGaTe$_2$ have shown a good thermoelectric figure of merit $ZT = 1.4$ realised experimentally \cite{Plirdpring} and the theoretical calculations also found a ZT of 1.69 \cite{VKG1}. Along the same line, we have calculated the thermoelectric properties of CuAlTe$_2$ and compared with the isostructural CuGaTe$_2$. We have calculated the thermopower and electrical conductivity scaled by relaxation time of CuAlTe$_2$ as a function of temperature for various levels of doping. The temperature dependent thermoelectric properties of CuAlTe$_2$ are presented in Fig. 5. As mentioned earlier, the anisotropic nature can be seen clearly from this figure along both $a$ and $c$-axes, and it increases with increase in temperature and doping concentration. We found almost similar thermopower for both electron and hole doping, though a bit higher in holes. However, $\sigma/\tau$ is found to be twice for electrons compared to holes (see Fig. 5). This confirms that the electron as charge carrier is more favourable than holes for CuAlTe$_2$. We compared the thermopower, electrical conductivity ($\sigma/\tau$) and electronic part of the thermal conductivity scaled by relaxation time (${\kappa_e}/\tau$) of CuAlTe$_2$ and CuGaTe$_2$ along both $a-$ and $c-$directions, and the same is shown in Fig. 6. From this figure we found that the variation of $\sigma/\tau$ and ${\kappa_e}/\tau$ for both the compounds along a-axis is almost similar, which might allows us to assume the relaxation time of CuAlTe$_2$ to be almost the same as that of CuGaTe$_2$ at 700 $K$, as it is well known that the electronic part of the thermal conductivity always dominate at high temperature. 

\par Further we also compared the thermopower, electrical conductivity scaled by relaxation time and power factor of CuAlTe$_2$ and CuGaTe$_2$ for both electron and hole carriers at $10^{19}$, $10^{20}$ $cm^{-3}$ concentrations (which is the optimum working range) at $700 K$, and the same is presented in Table-4. From the table it is quite evident that the power factor values of CuAlTe$_2$ are little higher compared to CuGaTe$_2$, except for electron doping along $a-$axis, and the difference is found to vanish at higher concentration and possess almost similar values at higher electron doping concentration. Though the electron doping is favourable in both the compounds, it conveys the impression that the power factor values for both electrons and holes are almost similar in the case of CuAlTe$_2$ especially for higher doping concentration (as seen from Table-4), which provokes us to state that CuAlTe$_2$ may find better device applications. To estimate the figure of merit for CuAlTe$_2$, we have taken the experimental thermal conductivity of CuAlTe$_2$ reported to be 3.108 $W/mk$ at a temperature of 700 $K$ \cite{Bodnar1}. We have also taken the relaxation time to be $2\times10^{-14} s$ at 700 $K$ as that of CuGaTe$_2$ \cite{VKG1}, since CuAlTe$_2$ and CuGaTe$_2$ are iso-structural to each other. With these values, we found the estimated figure of merit at 700 $K$ to be 1.88 and 1.93 for a concentration of $1\times10^{20} cm^{-3}$ holes and electrons respectively. This shows that both the carriers are beneficial, which provoke us to state CuAlTe$_2$ to be yet another promising TE material, which can lead to device applications. 
To confirm this we look forward for the experimentalist to explore more in this direction for the thermoelectric properties of Cu-based chalcopyrite materials.

\section{Conclusion}
Electronic structure and transport properties of CuAlCh$_2$, CuGaS$_2$ and CuGaSe$_2$ were calculated within the density functional theory. 
Electronic structure properties are calculated using the TB-mBJ functional and the obtained band gaps are found to be nearly close to the experimental results and very good in comparison with the other theoretical results. The calculated effective mass of the charge carriers show a combination of mixed light and heavy bands, which would be favourable for the thermoelectric properties. All the investigated compounds are found to be very good thermoelectric materials for electron doping. 
Among all the compounds CuAlTe$_2$ is found to be more promising candidate for both electron and hole doping. We compare the thermoelectric properties of CuAlTe$_2$ with the iso-structural CuGaTe$_2$, and the results predict almost similar TE properties of CuAlTe$_2$ in line with CuGaTe$_2$, which enable us to predict CuAlTe$_2$ to be yet another promising thermoelectric material. Our study on the Cu-based chalcopyrite semi-conductors has predicted CuAlTe$_2$ to be more promising with both electrons and holes as probable carriers which might enable device applications. We look forward for further experimental studies in this direction. 

\section{Acknowledgement}
V.K.G and V.K would like to acknowledge IIT-Hyderabad for providing computational facility. V.K.G. would like to thank MHRD for the fellowship. G.V thank Center for Modelling Simulation and Design-University of Hyderabad (CMSD-UoH) for providing computational facility.


\begin{table*}
\caption{Ground state properties of CuAlCh$_2$ ($\it{Ch}$ = S, Se, Te) and CuGaCh$_2$($\it{Ch}$=S, Se) with GGA functional along with the available experimental results.}
\begin{tabular}{lllllll}
\hline\noalign{\smallskip}
	&	&CuAlS$_2$	&CuAlSe$_2$	&CuAlTe$_2$	&CuGaS$_2$	&CuGaSe$_2$\\

\noalign{\smallskip}\hline\noalign{\smallskip}		
a(\AA)	&Present	&5.36	&5.68	&6.09	&5.38	&5.67	\\
	&Exp.$^a$	&5.334	&5.602 	&5.964	&5.356 	&5.614  \\
	
c(\AA)	&Present	&10.56	&11.17  &12.08	&10.17 &10.71\\
	&Exp.$^a$	&10.444 &10.944 &11.780 &10.435&11.03\\

u	&Present	&0.26	&0.25	&0.24	&0.25	&0.24\\
(Internal parameter of $Ch$) \\
        &Exp.$^a$	&0.275 	&0.269  &0.25	&0.275  &0.250  \\
\noalign{\smallskip}\hline
$^a$ Ref. \cite{Spiess} \\
\end{tabular}
\end{table*}

\begin{table*}
\caption{Calculated TB-mBJ band gaps including spin-orbit coupling compared with experimental and other theoretical values in eV.}
\begin{tabular}{ccccc}
\hline
		&Present	&Exp.	&Other theory \\
\hline
CuAlS$_2$	&2.56	&3.49$^a$	&2.7$^e$, 2.665$^f$\\

CuAlSe$_2$	&2.11	&2.65$^b$	&2.1$^e$, 2.446$^f$\\

CuAlTe$_2$	&1.86	&2.06$^c$	&1.6$^e$, 2.149$^f$\\

CuGaS$_2$	&1.71	&2.43$^d$	&1.416$^f$\\

CuGaSe$_2$	&1.04	&1.68$^d$	&1.416$^f$\\
\hline
$^a$ Ref. \cite{Shay},$^b$ Ref. \cite{Shirakata},$^c$ Ref. \cite{Bodnar},$^d$ Ref. \cite{Semi},$^e$ Ref. \cite{Reshak},$^f$ Ref. \cite{Raja}
\end{tabular}
\end{table*}

\begin{table*}
\caption{Calculated effective mass of Cu-based chalcopyrites in some selected directions of the Brillouin zone in the units of electron rest mass ($m_0$)}
\begin{tabular}{cccccccccccccccccc}
\hline
            	  	&CuAlS$_2$	&CuAlSe$_2$	&CuAlTe$_2$	&CuGaS$_2$	&CuGaSe$_2$	&CuGaTe$_2$\\
\hline
VBM\\
\hline
$\Gamma$-N$_{II}$ 	&2.24		&1.11		&0.44		&0.89		&0.56		&0.40\\

$\Gamma$-N$_I$ 		&3.84		&1.69		&0.63		&1.29		&0.46		&0.46\\

$\Gamma$-Z		&1.03		&0.53		&0.26		&0.37		&0.25		&0.16\\
\hline
CBM\\
\hline

$\Gamma$-N		&0.42		&0.21		&0.36		&0.26		&0.15		&0.16\\

$\Gamma$-Z		&0.83		&0.20		&1.98		&0.25		&0.15		&0.15\\	

\hline
\end{tabular}
\end{table*}

\begin{table*}
\caption{Comparison of thermopower, electrical conductivity scaled by relaxation time and power factor scaled by relaxation time of CuAlTe$_2$ and CuGaTe$_2$ for both electron and hole carriers at a doping concentration of $10^{19}$, $10^{20} cm^{-3}$ at a temperature of $700 K$}
\begin{tabular}{cccccccccccccccccc}
\hline
		&concentration $(cm^{-3})$			&	&10$^{19}$	&	&	&	&10$^{20}$\\
\hline
		&carriers 					&n$_h$	&	&n$_e$	&	&n$_h$	&	&n$_e$	&	\\
\hline
		&Direction					&a	&c	&a	&c	&a	&c	&a	&c	\\
\hline\\
CuAlTe$_2$	&S($\mu V/K$)					&398.00	&358.61	&360.92	&386.13	&281.38	&183.72	&201.26	&234.87	\\

		&($\sigma/\tau)x10^{18}(\Omega ms)^{-1}$	&0.58	&0.44	&1.19	&0.33	&5.27	&3.75	&10.6	&3.06	\\
		
		&(S$^2\sigma/\tau)x10^{11}(W/mK^2s)$		&0.92	&0.57	&1.55	&0.49	&4.17	&1.27	&4.29	&1.69	\\		

\hline\\
CuGaTe$_2$	&S($\mu V/K$)					&359.87	&336.77	&347.27	&267.09	&175	&167.65	&178.15	&114.70	\\

		&($\sigma/\tau)x10^{18}(\Omega ms)^{-1}$	&0.68	&0.20	&1.59	&0.55	&6.75	&1.74	&13.6	&3.06	\\
		
		&(S$^2\sigma/\tau)x10^{11}(W/mK^2s)$		&0.88	&0.23	&1.91	&0.39	&2.06	&0.49	&4.32	&0.40	\\		
\hline
\end{tabular}
\end{table*}

\begin{figure*}
\begin{center}
\subfigure[]{\includegraphics[width=65mm,height=80mm]{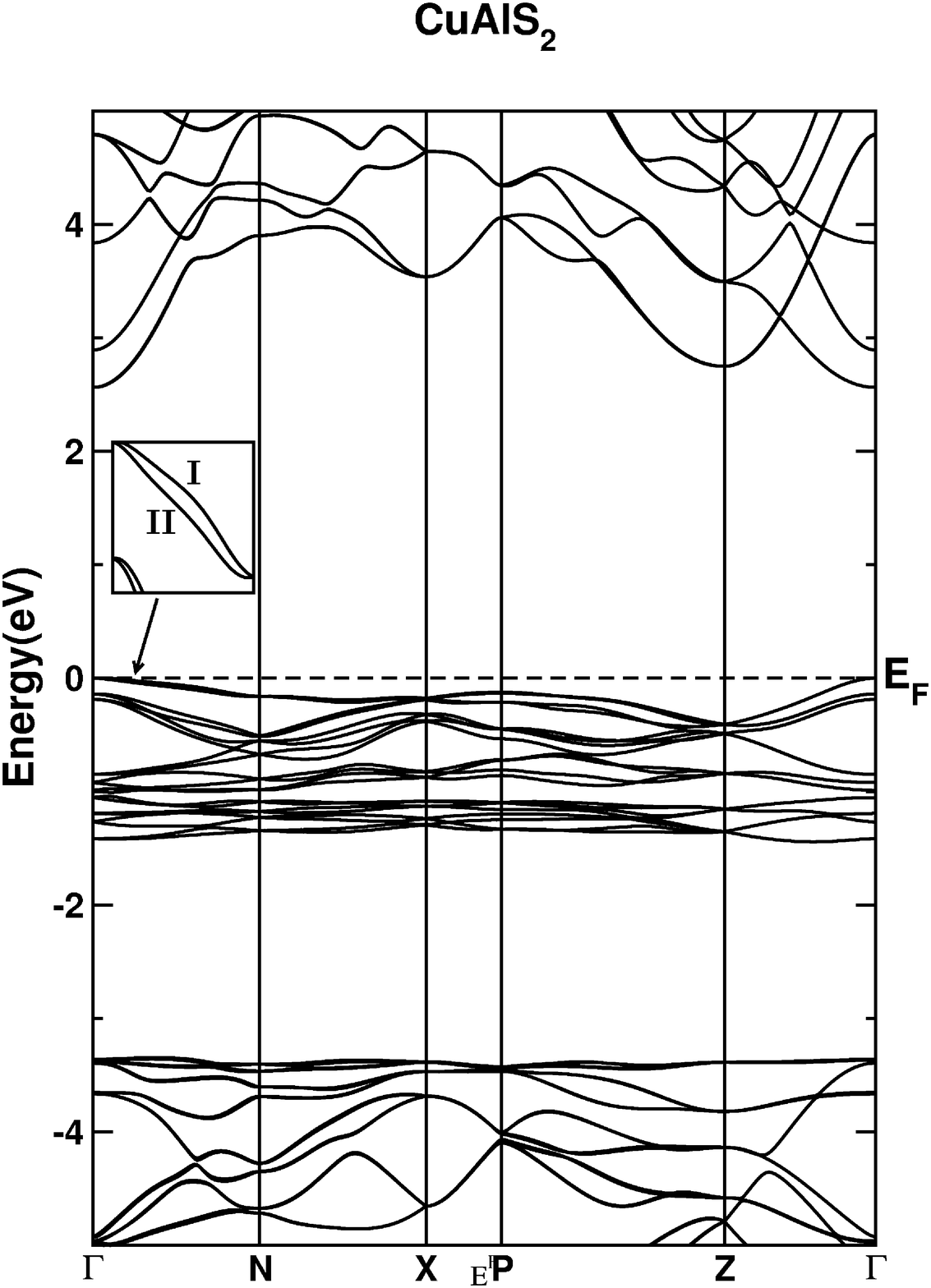}}
\subfigure[]{\includegraphics[width=65mm,height=65mm]{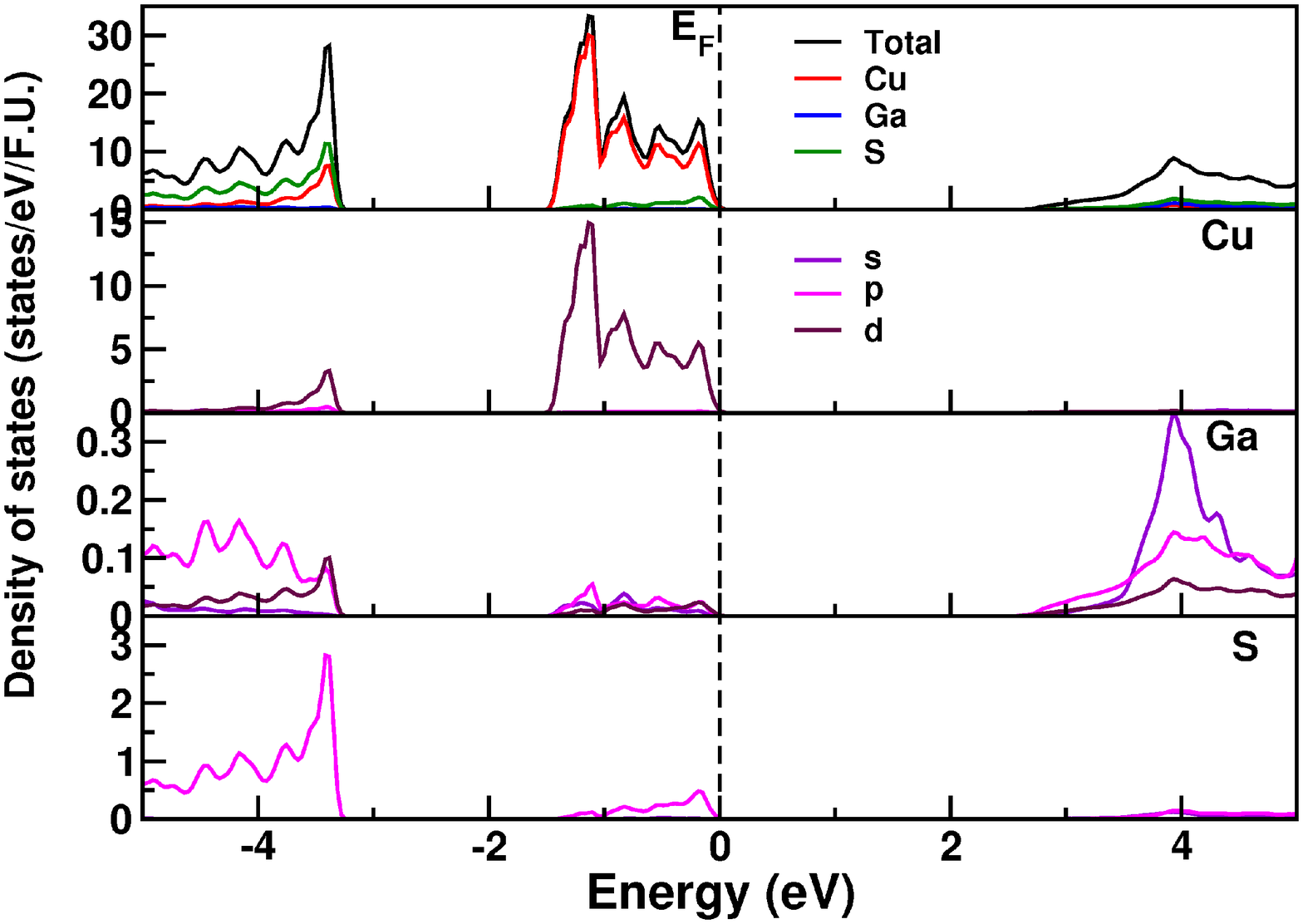}}
\caption{(Color online) Calculated (a) Band structure and (b) Density of states of CuAlS$_2$. The energy zero corresponds to the VBM.}
\end{center}
\end{figure*}

\begin{figure*}
\begin{center}
\includegraphics[width=120mm,height=70mm]{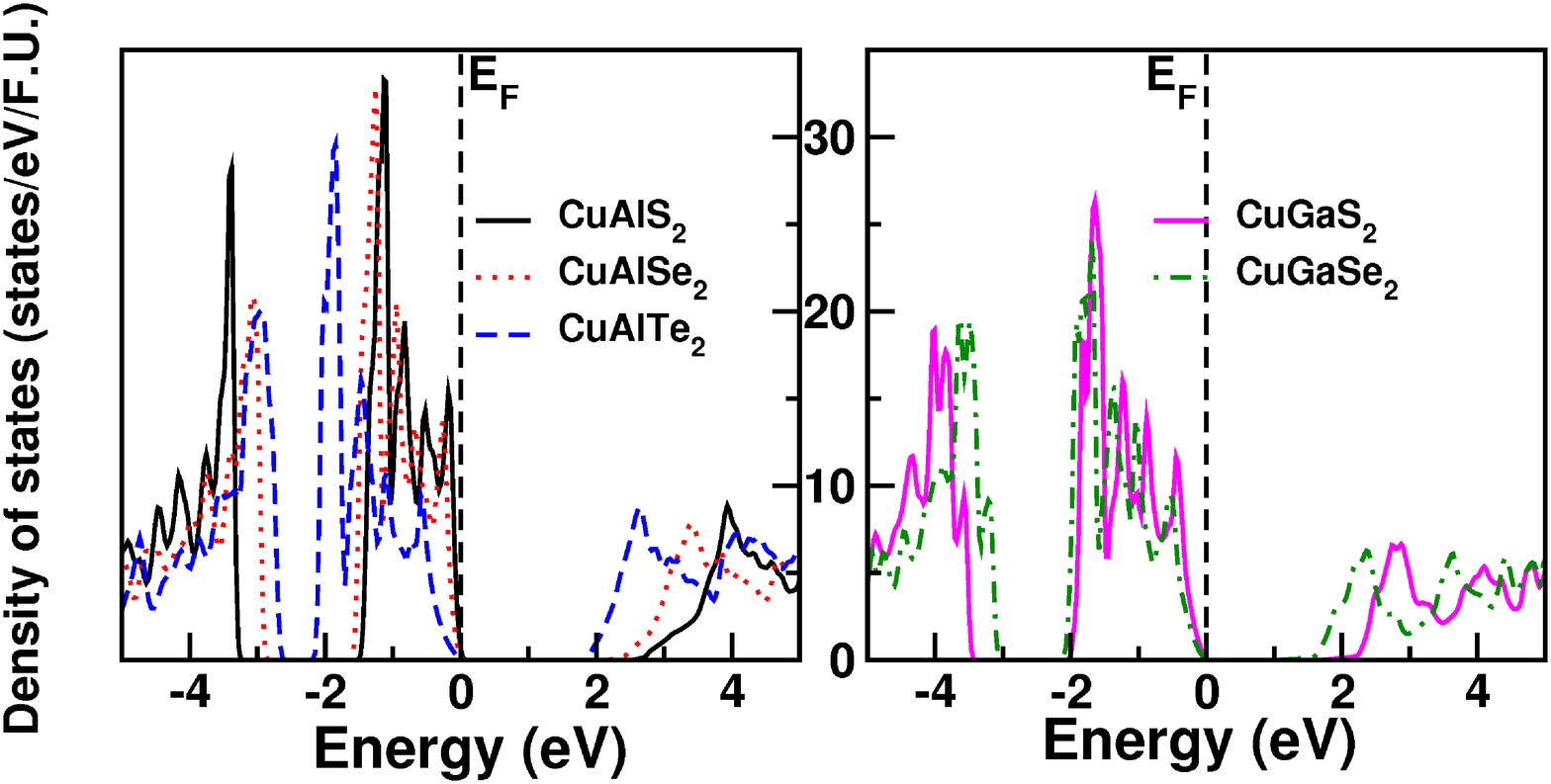}
\caption{(Color online) Calculated Density of states of CuXCh$_2$. The energy zero corresponds to the VBM.}
\end{center}
\end{figure*}


\begin{figure*}
\begin{center}
\includegraphics[width=130mm,height=120mm]{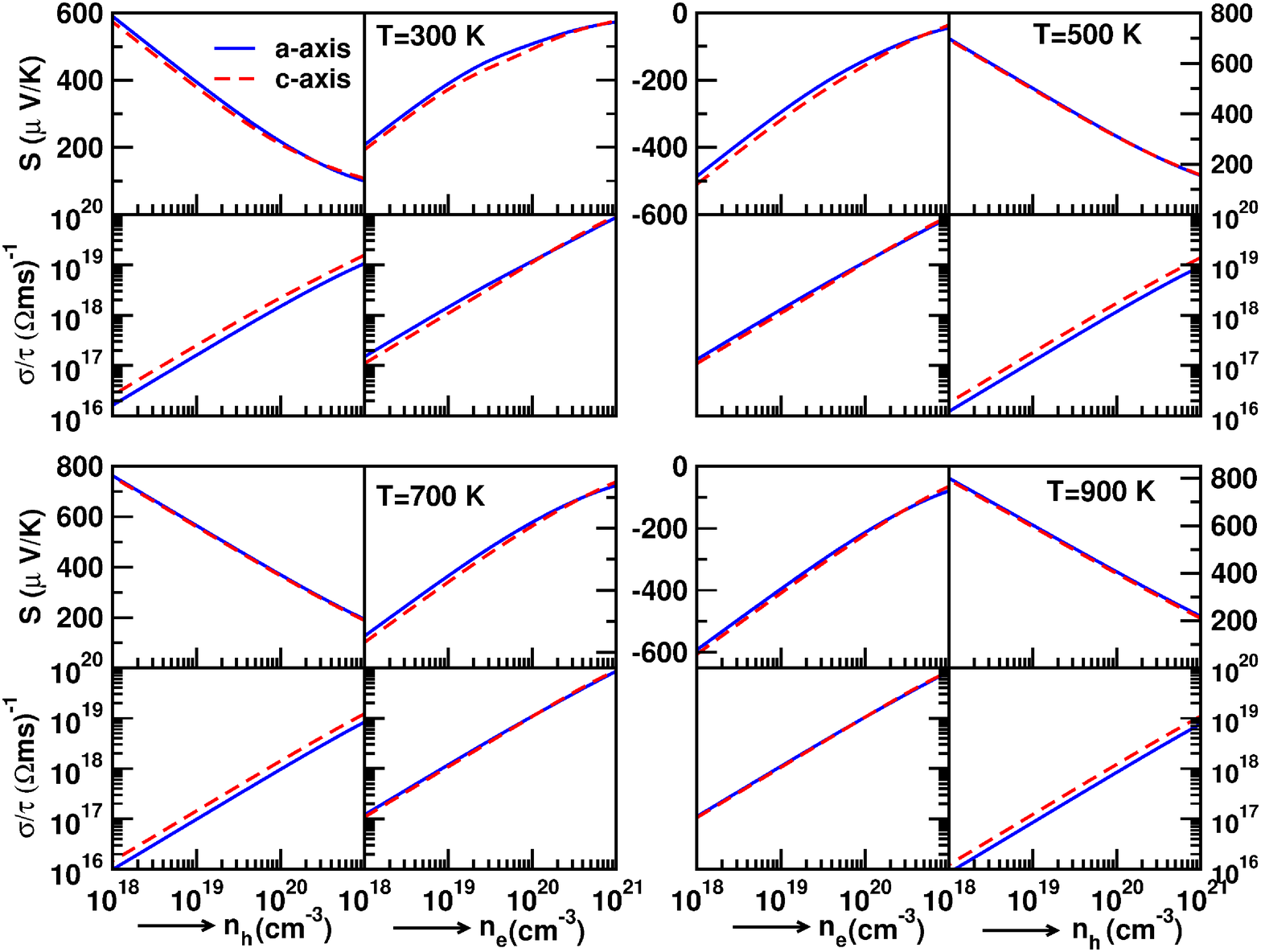}
\caption{(Color online) Variation of thermoelectric properties of CuAlS$_2$ as a function of temperature}
\end{center}
\end{figure*}


\begin{figure*}
\begin{center}
\includegraphics[width=120mm,height=120mm]{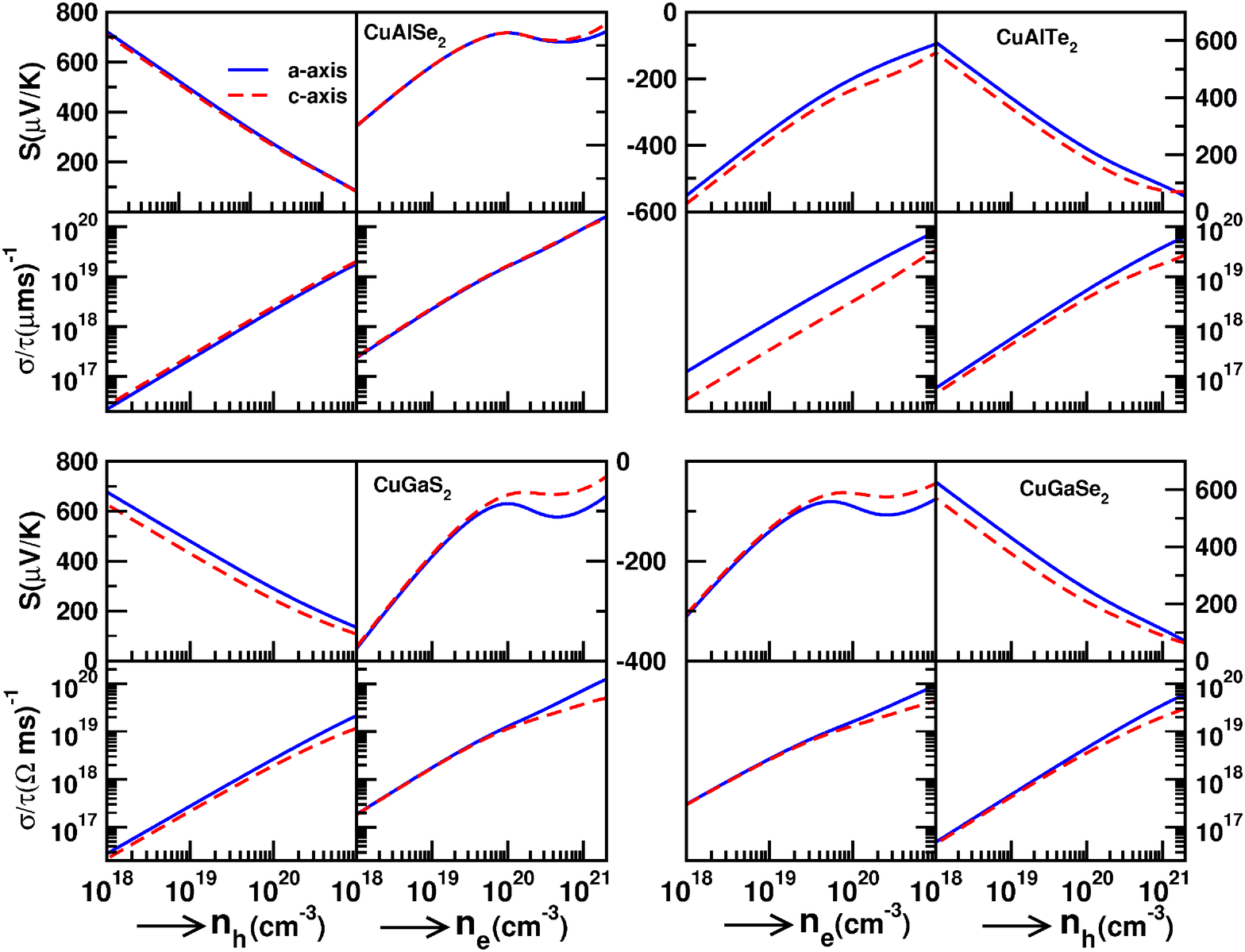}
\caption{(Color online) Variation of thermoelectric properties of CuAlCh$_2$ (Ch = Se, Te) and CuGaCh$_2$ (Ch = S, Se) at 700 $K$}
\end{center}
\end{figure*}


\begin{figure*}
\begin{center}
\includegraphics[width=110mm,height=110mm]{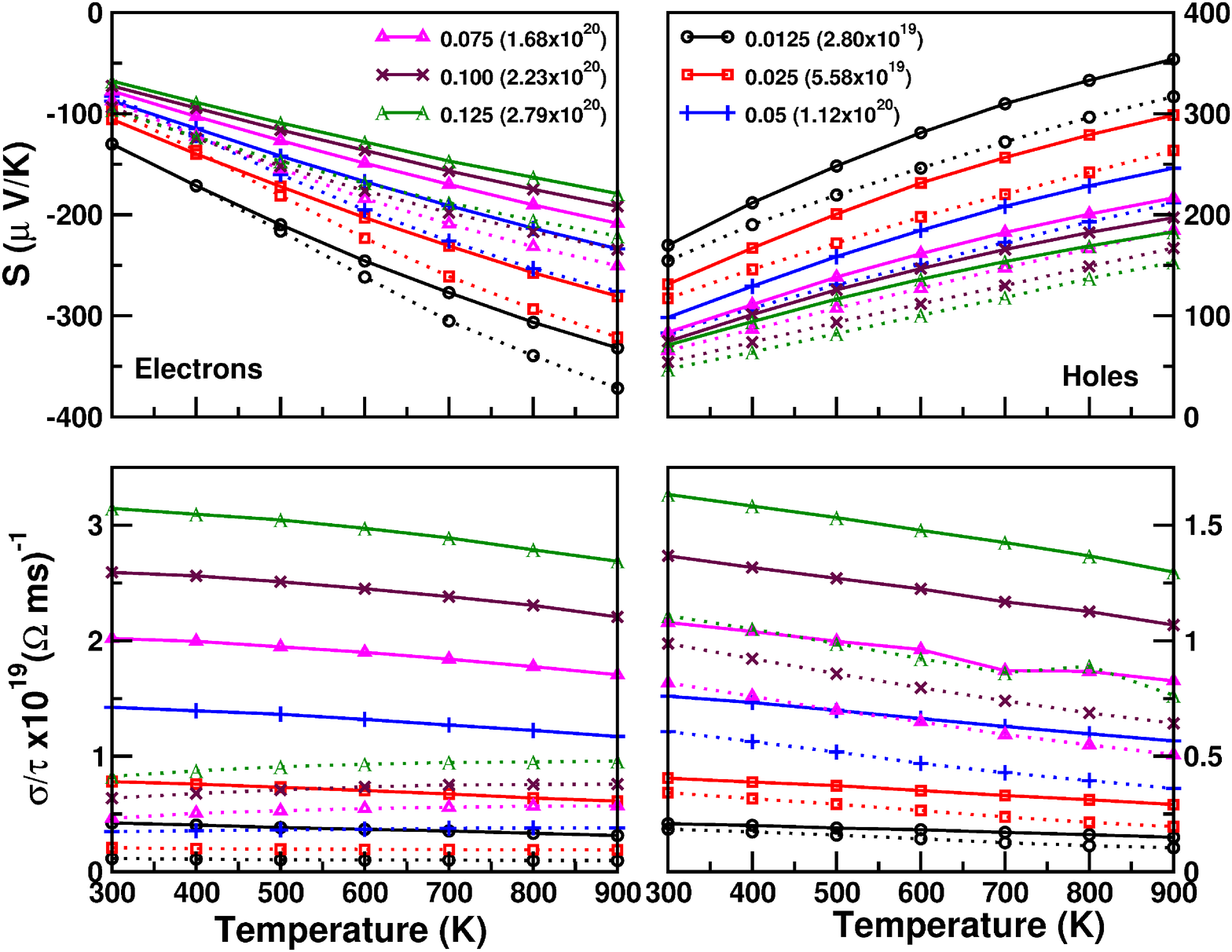}
\caption{(Color online) Thermopower and electrical conductivity for CuAlTe$_2$ for various hole (right) and electron (left) concentrations along the a-axis (solid line with symbols) and c-axis (dotted line with symbols)}
\end{center}
\end{figure*}

\begin{figure*}
\begin{center}
\subfigure[]{\includegraphics[width=100mm,height=60mm]{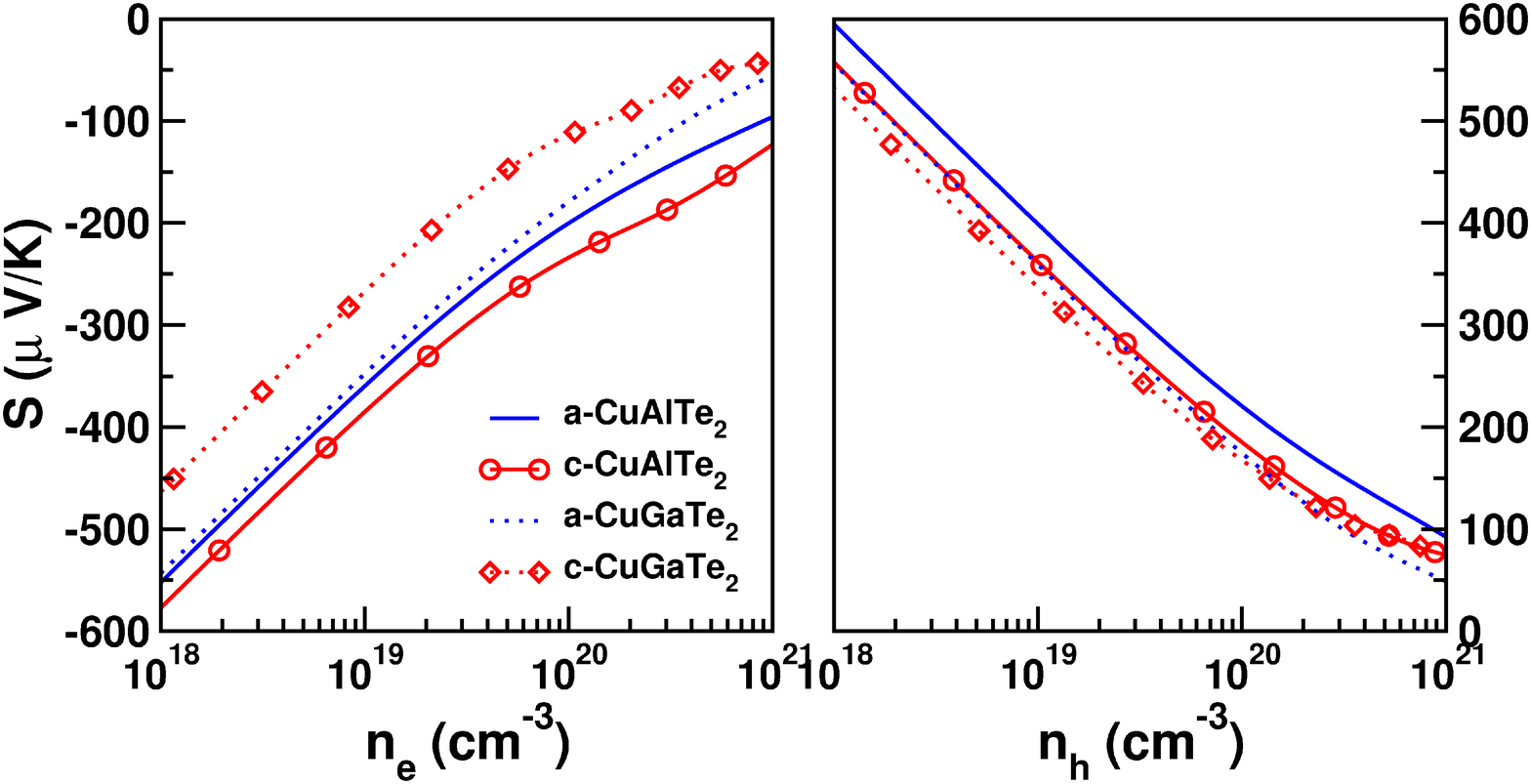}}\\
\subfigure[]{\includegraphics[width=100mm,height=60mm]{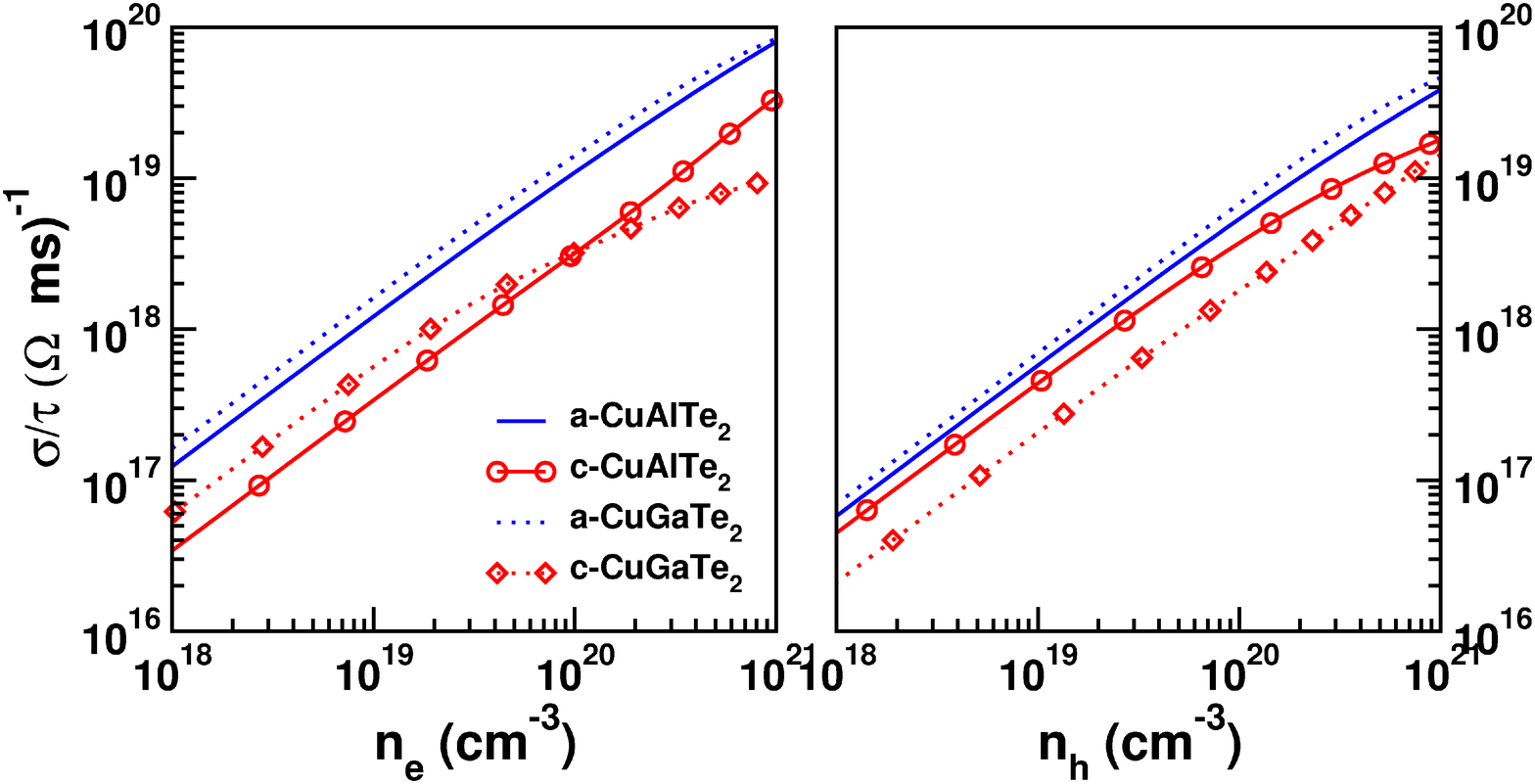}}\\
\subfigure[]{\includegraphics[width=100mm,height=60mm]{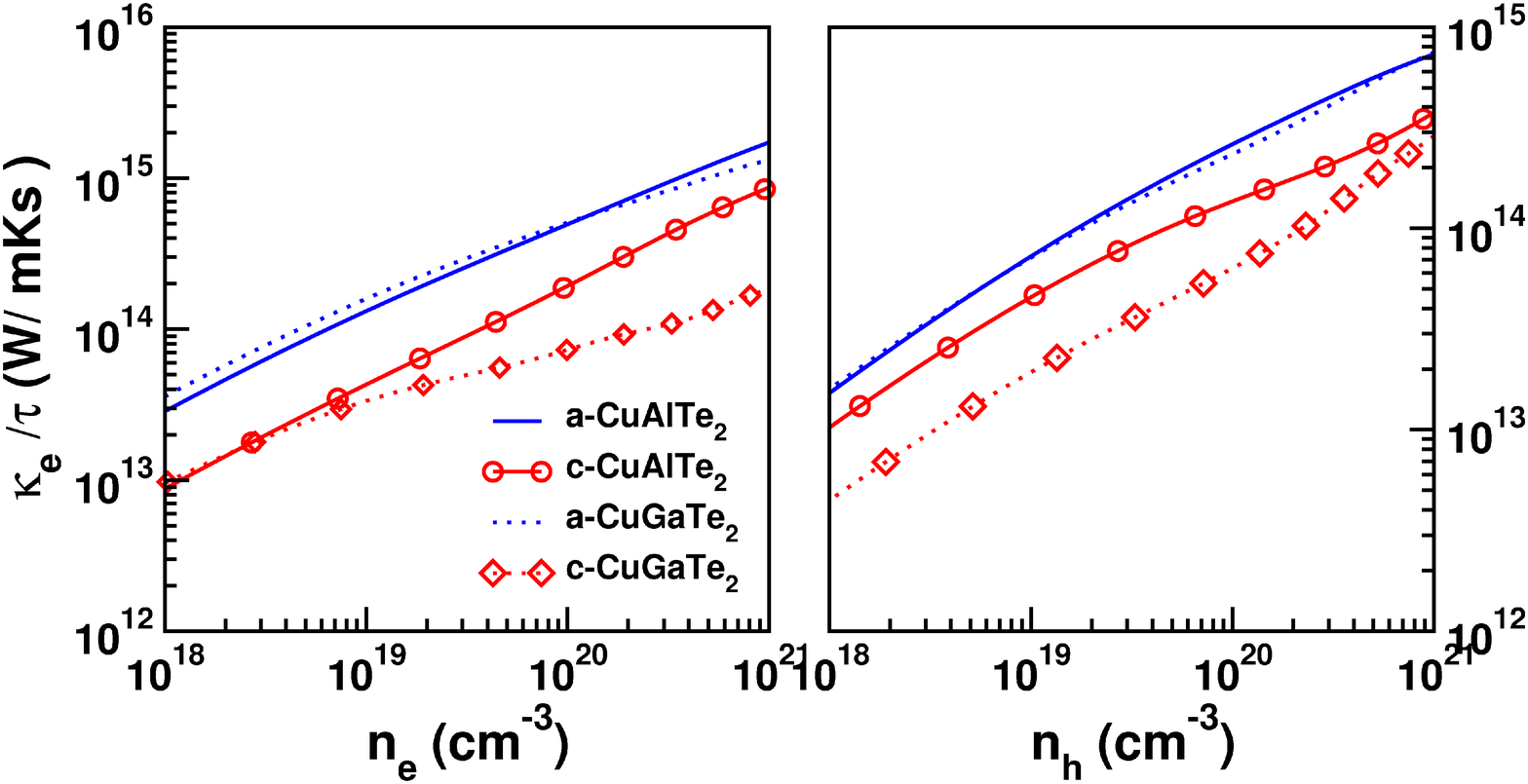}}
\caption{(Color online) Comparison of (a) thermopower (b) electrical conductivity and (c) electronic part of the thermal conductivity for CuAlTe$_2$ and CuGaTe$_2$ for holes (right) and electrons (left) at a temperature of 700 $K$}
\end{center}
\end{figure*}

\end{document}